\documentclass[aps,prl,twocolumn,superscriptaddress,showpacs,amsmath,amssymb]{revtex4-1}
\usepackage{amsmath}
\usepackage{amsfonts}
\usepackage{amssymb}
\usepackage{graphicx}
\usepackage{color}

\usepackage{bm}
\usepackage{braket}
\usepackage{hyperref}
\usepackage{wasysym}
\usepackage{stmaryrd}

\begin{document}

\title{Quantum liquid crystals in the finite-field K$\Gamma$ model for $\alpha$-RuCl$_3$}

\author{Masahiko G. Yamada}
\email[]{myamada@mp.es.osaka-u.ac.jp}
\affiliation{Department of Materials Engineering Science, Osaka University, Toyonaka 560-8531, Japan}
\author{Satoshi Fujimoto}
\affiliation{Department of Materials Engineering Science, Osaka University, Toyonaka 560-8531, Japan}
\affiliation{Center for Quantum Information and Quantum Biology, Osaka University, Toyonaka 560-8531, Japan}

\date{\today}

\begin{abstract}
We study the extended Kitaev model called the K$\Gamma$ model,
using a perturbative expansion combined with
a well-controlled mean-field approximation and a cutting-edge
exact diagonalization.  In the phase diagram,
we discover a nematic Kitaev spin liquid and a Kekul\'e Kitaev
spin liquid.  The former potentially
explains the high-field nematic state with zero Chern number
experimentally observed in $\alpha$-RuCl$_3$,
even for sufficiently small values of $\Gamma/|K|$.
The latter has a Majorana zero mode in its $Z_3$
vortex core, which can be potentially controlled by the domain
wall motion.  This opens a possible application of the quantum
liquid crystal phases in the K$\Gamma$ model to the topological
quantum computation.
\end{abstract}

\maketitle

\textit{Introduction}. ---
The Kitaev model~\cite{Kitaev2006} is a prominent example of exactly solvable models with
a quantum spin liquid (QSL) property, which is characterized by various topological
signatures~\cite{Obrien2016,Yamada2021topological}.  However, it is known that real
materials cannot fully be described by this fine-tuned model and include
additional interactions~\cite{Wang2020}, which potentially change the topological
property of the ground state.  Recently the K$\Gamma$ model~\cite{Rau2014,Gohlke2018,Catuneanu2018,ZXLiu2018,Samarakoon2018,Gordon2019,Chern2020,Lee2020,TYamada2020,Gohlke2020,Buessen2021,Luo2021,Zhang2021}
has been investigated intensively in connection with a Kitaev
material $\alpha$-RuCl$_3$, and various kinds of topological
transitions were expected theoretically.  Whereas
various numerical methods have been used, most studies treat
itinerant Majorana fermions of the Kitaev model indirectly,
and the role of Majorana fermions is unclear.

Among many possible realizations of the proximate Kitaev model~\cite{Jackeli2009,Plumb2014,Yamada2017,Yamada2017xsl,Liu2018,Sano2018,Jang2019},
$\alpha$-RuCl$_3$ is an outstanding material because a half-integer
quantization of the thermal Hall conductivity has been observed~\cite{Kasahara2018,Yokoi2020}.
At higher field, the breaking of the threefold rotation symmetry
has been observed from the field angle dependence of the heat capacity
at the same time as the disappearance of the thermal Hall effect~\cite{Tanaka2020}.
The coincidence is attributed to the first-order transition from
Kitaev's $B$ phase with a Chern number 1 to Kitaev's $A$ phase with
a Chern number 0, which occurs directly maintaining an energy
gap~\cite{Takahashi2021}.  A theoretical characterization is necessary for the high-field
nematic phase based on a microscopic model like the K$\Gamma$ model.
While some studies~\cite{Gohlke2018,Lee2020}
find the nematic phase within the K$\Gamma$ model, the property
of the associated nematic phase transition from the Kitaev spin liquid
is not well discussed~\cite{Gohlke2020}.

Another interesting phase with a broken threefold rotation symmetry
is a Kekul\'e phase.  Though both a nematic phase and a Kekul\'e phase
are $Z_2$ spin liquids with a toric code topological order,
some properties are different.  The most significant difference
is the presence of a Majorana zero mode (MZM) in the $Z_3$ vortex for a
Kekul\'e phase~\cite{Jackiw1981,Yang2019}.  Here
a $Z_3$ vortex means a crossing point of boundaries of three different domains
of the $Z_3$-broken Kekul\'e phases.  MZMs can potentially be braided
by the domain wall motion, which leads to a nontrivial phase.
Thus, by controlling such topological defects in this liquid crystalline
phase, the protected quantum computation in the MZM states is possible.

Both of these phases are relevant to the K$\Gamma$ model even in the
mean-field level, and thus we investigate these phases precisely.
From now on we call the nematic phase nematic Kitaev spin liquid (NKSL)
and the Kekul\'e phase Kekul\'e-Kitaev spin liquid (KKSL).

From the mean-field approximation and the exact diagonalization,
we discover a vast region of NKSL and KKSL with a broken threefold
rotation symmetry.  Both of these phases can be regarded as quantum spin
analogues of liquid crystals~\cite{Kivelson1998}.
These phases are enabled to be discovered
by the methods which treat itinerant Majorana fermions directly.

In this Letter, we study a phase diagram of the finite-field
K$\Gamma$ model using the third-order perturbation.  The effective
model is an interacting Majorana model with four-body and six-body
interactions~\cite{Bravyi2010,Vijay2015,Affleck2017,Li2018,Kamiya2018,Wamer2018,Rahmani2019,Rahmani2019review,Li2019,Tummuru2021}.
Based on the mean-field solution and the exact diagonalization,
we find the phase diagram is rich enough to predict
various exotic spin liquids like NKSL and KKSL.

\begin{figure}
\centering
\includegraphics[width=8.6cm]{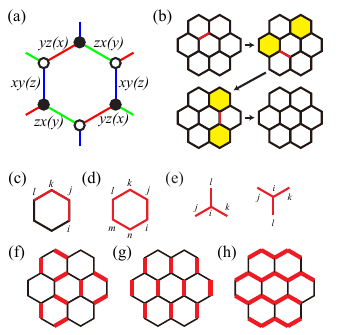}
\caption{(a) Honeycomb lattice where the K$\Gamma$ model is defined.
Red, green, and blue bonds represent bonds in the $x$-, $y$-, and
$z$-directions, respectively.
(b) Example of intermediate states with an energy $0.227|K|/4$ in the
third-order perturbation.
(c) Example of armchair-type four-body interactions shown in red bonds.
All six symmetry-equivalent terms are included in the Hamiltonian.
(d) Six-body plaquette interaction shown in red.
(e) $\Yup$-shaped and $\Ydown$-shaped four-body interactions shown in red.
(f) Kekul\'e bond order with strong bonds shown in red.
(g) Nematic bond order with strong bonds shown in red.
(h) Zigzag nematic bond order with strong bonds shown in red.}
\label{honeycomb}
\end{figure}

\textit{Third-order perturbation}. ---
We begin with the following Kitaev-$\Gamma$ or K$\Gamma$ model.
These two terms are known to be dominant in $\alpha$-RuCl$_3$.
\begin{align}
    H_0 &= K\sum_{\langle ij \rangle \in \alpha \beta (\gamma)} S_i^\gamma S_j^\gamma + \Gamma \sum_{\langle ij \rangle \in \alpha \beta (\gamma)} (S_i^\alpha S_j^\beta + S_i^\beta S_j^\alpha) \nonumber \\
    &= -\frac{|K|}{4}\sum_{\langle ij \rangle \in \alpha \beta (\gamma)} \sigma_i^\gamma \sigma_j^\gamma + \frac{\Gamma}{4} \sum_{\langle ij \rangle \in \alpha \beta (\gamma)} (\sigma_i^\alpha \sigma_j^\beta + \sigma_i^\beta \sigma_j^\alpha),
\end{align}
where we assume $K<0$, and $\Gamma$ to be a real number,
and $S_i$ and $\sigma_i$ are spin-1/2 and Pauli operators
defined on the $i$th site, respectively.
$\langle ij \rangle \in \alpha \beta (\gamma)$ means that a
nearest-neighbor (NN) bond $\langle ij \rangle$ belongs to
$\gamma$-bonds as shown in Fig.~\ref{honeycomb}(a),
and the $\alpha \beta$-plane is perpendicular to the
$\gamma$-direction.
We note that our study is based on the original Kitaev model
with an additional interaction, not the so-called Kekul\'e-Kitaev model~\cite{Kamfor2010,Quinn2015,Mirmojarabian2020}.
The Kekul\'e-Kitaev model explicitly breaks the $Z_3$ symmetry, while
our model spontaneously breaks this symmetry in the Kekul\'e phase.

As is usually the case, the nontrivial contribution begins
from the third order in $\Gamma/K$.  The third-order
perturbation leads to the following effective Hamiltonian.
\begin{align}
    H_\textrm{eff} =& \frac{6\Gamma^3}{0.21 |K|^2} \sum_{\langle ij \rangle \langle kl \rangle \langle mn \rangle \in \hexagon} (\sigma_i^x \sigma_j^y + \sigma_i^y \sigma_j^x) \nonumber \\
    &\times (\sigma_k^y \sigma_l^z + \sigma_k^z \sigma_l^y)(\sigma_m^z \sigma_n^x + \sigma_m^x \sigma_n^z),
\end{align}
where $\langle ij \rangle \langle kl \rangle \langle mn \rangle \in \hexagon$ means that
$\langle ij \rangle$, $\langle kl \rangle$, and $\langle mn \rangle$ are two possible patterns
for $z$-, $x$-, and $y$-bonds, respectively, inside each hexagon.
The factor $0.21 \sim 4\times 0.227^2$ comes from the intermediate
vortex pattern~\cite{Yamada2020,Takahashi2021}, as shown in Fig.~\ref{honeycomb}(b).

We then move on to the effective Majorana interacting model.
\begin{align}
    H_c =& \frac{t}{4}\sum_{\langle ij \rangle} ic_i c_j -\frac{6\Gamma^3}{0.21 |K|^2}\sum_{\langle ijkl \rangle} c_i c_j c_k c_l \nonumber \\
    &-\frac{12\Gamma^3}{0.21 |K|^2} \sum_{\langle ijklmn \rangle} ic_i c_j c_k c_l c_m c_n,
\end{align}
where $t \propto |K| + O(\Gamma^2/|K|)$, $\langle ijkl \rangle$
represents armchair-type interactions shown in Fig.~\ref{honeycomb}(c),
and $\langle ijklmn \rangle$ represents six-body interactions
shown in Fig.~\ref{honeycomb}(d).  The site ordering of
$\langle ijkl \rangle$ and $\langle ijklmn \rangle$ is
counterclockwise, as shown in Figs.~\ref{honeycomb}(c)-(d).

The same process applies to the case with a magnetic field.
We can consider the Hamiltonian $H = H_0 + H_1$ with
\begin{align}
    H_1 &= -\sum_j (h^x S_j^x + h^y S_j^y + h^z S_j^z) \nonumber \\
    &= -\sum_j \left(\frac{h^x}{2} \sigma_j^x + \frac{h^y}{2} \sigma_j^y + \frac{h^z}{2} \sigma_j^z \right),
\end{align}
where $\vec{h} = (h^x, h^y, h^z)^t$ is an applied magnetic field.
By tracking intermediate vortex states, we would finally get
\begin{align}
    H_c^\prime =& \frac{t}{4}\sum_{\langle ij \rangle} ic_i c_j - \frac{12\Gamma^3}{0.21 |K|^2} \sum_{\langle ijklmn \rangle} ic_i c_j c_k c_l c_m c_n \nonumber \\
    &- \sum_{\langle ijkl \rangle \in \alpha\beta(\gamma)} \left(\frac{6h^\alpha h^\beta \Gamma}{0.060|K|^2} + \frac{6\Gamma^3}{0.21 |K|^2} \right) c_i c_j c_k c_l \nonumber \\
    &+ \sum_{\langle\!\langle\!\langle ij \rangle\!\rangle\!\rangle \in \alpha\beta(\gamma)} \frac{12h^\alpha h^\beta \Gamma}{0.060|K|^2} ic_ic_j +\frac{6h^x h^y h^z}{0.035|K|^2} \sum_{\langle\!\langle ij \rangle\!\rangle}ic_i c_j \nonumber \\
    & + \frac{6h^x h^y h^z}{0.035|K|^2} \left(\sum_{\langle\!\langle ijkl \rangle\!\rangle_\Yup} c_i c_j c_k c_l - \sum_{\langle\!\langle ijkl \rangle\!\rangle_\Ydown} c_i c_j c_k c_l \right), \label{eff}
\end{align}
where $t \propto |K| + O(\Gamma^2/|K|) + O(h^2/|K|)$,
$\langle ijkl \rangle \in \alpha\beta(\gamma)$ means
that a bond $\langle jk \rangle$ belongs to $\gamma$-bonds
with the $\alpha\beta$-plane perpendicular to the $\gamma$-direction,
$\langle\!\langle ij \rangle\!\rangle$ represents a
next-nearest-neighbor (NNN) bond, and
$\langle\!\langle ijkl \rangle\!\rangle$ represents
a $\Yup$-shaped, or $\Ydown$-shaped interaction for sites $ijkl$,
where $ij$, $jk$, and $il$ are connected by $x$-, $y$-, and $z$-bonds,
respectively, as shown in Fig~\ref{honeycomb}(e).
$\langle\!\langle\!\langle ij \rangle\!\rangle\!\rangle \in \alpha\beta(\gamma)$
means the next-next-nearest-neighbor (NNNN) bond which is a diagonal
of hexagons in the $\gamma$-direction, where $i$ is even and $j$ is
odd as usual.

From now on, we set $h^x = h^y = h^z = h/\sqrt{3}$ for simplicity.
The absolute value of $t$ is undecidable, so we set $t=1$ as a simple
normalization.  We note that two-body terms are not antisymmetrized,
so the hopping amplitude is $1/8$ in the Hermitian form.

\textit{The $tqs$ model}. ---
Let us begin with the zero-field model.  First, we extend the model
to the following form:
\begin{align}
    H_{tqs} =& \frac{t}{4}\sum_{\langle ij \rangle} ic_i c_j -q\sum_{\langle ijkl \rangle} c_i c_j c_k c_l \nonumber \\
    &-2s \sum_{\langle ijklmn \rangle} ic_i c_j c_k c_l c_m c_n,
\end{align}
with $q$ and $s$ being real parameters.
When $q = s = 6\Gamma^3/(0.21 |K|^2)$, this $tqs$ model becomes
the original (K$\Gamma$) model with $h = 0$.

The operator
\begin{equation}
    V_p = -ic_i c_j c_k c_l c_m c_n.
\end{equation}
can be defined for each hexagon plaquette $p=\langle ijklmn \rangle$,
and plays an important role as they commute with each other.
Indeed, the $tqs$ model is integrable in the limit $|s| \to \infty$.
Eigenstates are constructed by assigning $V_p = \pm 1$ for each
plaquette.

\begin{figure}
    \centering
    \includegraphics[width=8.6cm]{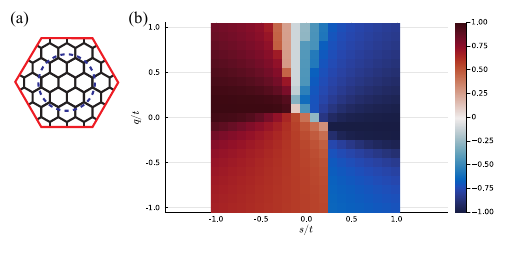}
    \caption{(a) 54-site cluster with a periodic boundary
    condition used in the exact diagonalization.
    A blue dashed circle shows an entanglement cut for
    the calculation of the entanglement entropy.
    (b) Expectation value of $V_p$ of the $tqs$ model
    computed by exact diagonalization.}
    \label{tqs}
\end{figure}

This limit ($|s| \to \infty$) is also known as the Vijay-Hsieh-Fu
(VHF) surface code~\cite{Vijay2015} and has a characteristic $Z_2$ topological order
with an exact $S_3$ anyon symmetry.  To check the existence of
such a phase we did exact diagonalization of the 54-site
cluster, which is shown in Fig.~\ref{tqs}(a).
This VHF surface code is indeed realized in the $tqs$ model
with a large $|s|$ region, as shown in Fig.~\ref{tqs}(b).
In Fig.~\ref{tqs}(b) the right half is mostly blue connected to the
$s\to \infty$ limit, and the left half is mostly red connected to
the $s\to \infty$ limit.  This means that the large $|s|$ phase is
rather stable with respect to the perturbation of $t$ and $q$.

It seems that even on the $q = s$ line, most of the $q>0$ region is
connected to this VHF surface code phase, and from this we conclude
that even in the original (K$\Gamma$) model Eq.~\eqref{eff} this phase is realized
in the large $\Gamma/|K|$ region.  Since the resolution of
Fig.~\ref{tqs}(b) is not good enough, the exact region of $\Gamma/|K|$
where this phase is realized will be discussed later.

\textit{K$\Gamma$ model with a magnetic field}. ---
Then, we go back to the original (K$\Gamma$) model Eq.~\eqref{eff}.
With a magnetic field, it indeed becomes easy to identify the phase
because we can use a Chern number to diagnose gapped topological
phases.  The Chern number is connected to the topological order
of the spin model according to Kitaev's 16-fold way.  For example,
phases with a Chern number $\pm 1$ in the Majorana model can
be identified with the Ising topological order, which we call
a chiral spin liquid (CSL) with a Chern number $\pm 1$ in the
following.

\begin{figure}
    \centering
    \includegraphics[width=8cm]{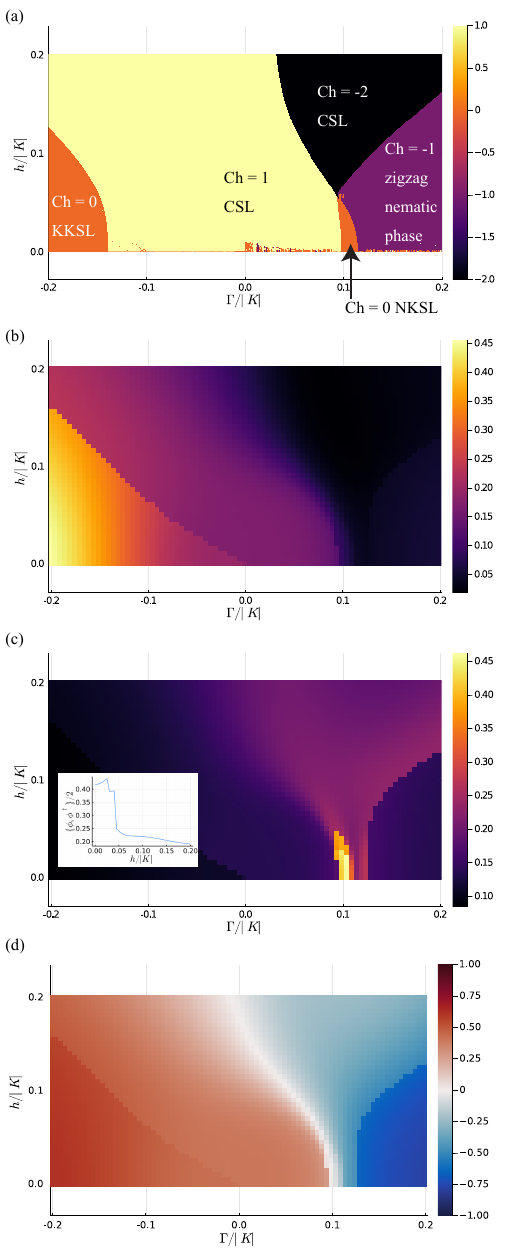}
    \caption{(a) Mean-field phase diagram of the K$\Gamma$ model.
    The color plot shows the value of the Chern number (Ch).
    (b) Hermitian version of a Kekul\'e order parameter
    $\{\psi, \psi^\dagger \}/2$ computed by exact
    diagonalization.  (c) Hermitian version of a
    nematic order parameter $\{\phi, \phi^\dagger \}/2$
    computed by exact diagonalization.  The inset shows
    $\{\phi, \phi^\dagger \}/2$ on the line $h/|K|=0.1$.  The value
    clearly drops from around 0.4 to 0.25 at the transition point.
    (d) $V_p$ computed by exact diagonalization.}
    \label{phase}
\end{figure}

The mean-field phase diagram is shown in Fig.~\ref{phase}(a)
Phases are identified from the Chern number and bond order
parameters.  The Chern number is computed by the
Fukui-Hatsugai-Suzuki method~\cite{Fukui2005}.  The details of the calculation is
included in Supplemental Material (SM)~\cite{SM}.  The method
is based on Refs.~\cite{Li2018,Takahashi2021}. From the phase
diagram we find the following phases:
\begin{enumerate}
    \item a KKSL with a Chern number $0$.
    \item Kitaev's non-Abelian CSL with a Chern number $1$.
    \item an Abelian CSL with a Chern number $-2$.
    \item an NKSL with a Chern number $0$.
    \item a zigzag nematic phase with a Chern number $-1$.
\end{enumerate}
The bond ordering patterns for a KKSL, an NKSL, and a zigzag nematic
phase are shown in Figs.~\ref{honeycomb}(f), (g), and (h), respectively.

\textit{Exact diagonalization}. ---
From now on we will confirm the results of the mean-field theory
by exact diagonalization of Eq.~\eqref{eff}.  The exact diagonalization
is done using the 54-site cluster again.  We note that this size is much
larger than what has been used for the Kitaev model exact diagonalization.
We take an average of expectation values for all the degenerate
ground states.  First, we check the operator
$\{\psi,\psi^\dagger\}/2$, corresponding to the absolute value
of the Kekul\'e order parameter~\cite{Pujari2015} with
\begin{equation}
    \psi = \Delta_1^\prime + e^{2\pi i/3} \Delta_2^\prime + e^{4\pi i/3} \Delta_3^\prime,
\end{equation}
where the definition of $\Delta_i^\prime$ ($i = 1$--$9$) is included in
SM~\cite{SM}.  In Fig.~\ref{phase}(b),
the orange triangular region with $\Gamma < 0$ corresponds to the
Kekul\'e-ordered phase.  The area of this phase becomes larger than
that of the mean-field phase diagram.  Surprisingly, it looks like
the Kekul\'e order appears with an infinitesimal $\Gamma < 0$ in
the 54-site calculation.  This implies that the Kitaev spin liquid
is unstable with respect to a negative $\Gamma$ perturbatively
and it is possible that the Kekul\'e order opens a gap for Dirac
cones (see SM~\cite{SM} for more details).

Next we check the nematic order parameter $\{\phi,\phi^\dagger\}/2$~\cite{Pujari2015}
with
\begin{equation}
    \psi = \Delta_1 + e^{2\pi i/3} \Delta_2 + e^{4\pi i/3} \Delta_3,
\end{equation}
where the definition of $\Delta_i$ ($i = 1$--$9$) is included in
SM~\cite{SM}.  In Fig.~\ref{phase}(c), NKSL is identified with
the small orange region around $\Gamma/|K| = 0.1$.  The area of this phase
becomes smaller than that of the mean-field phase diagram.  We now
confirmed the existence of NKSL even in the exact diagonalization,
but the zigzag nematic phase disappears in the same calculations.
There remains a violet region which is different from the zigzag
nematic phase found in the mean-field approach.
We note that the order parameter is nonzero
on the whole parameter space in Fig.~\ref{phase}(b) and (c)
due to the finite-size effect.

\begin{figure}
    \centering
    \includegraphics[width=8cm]{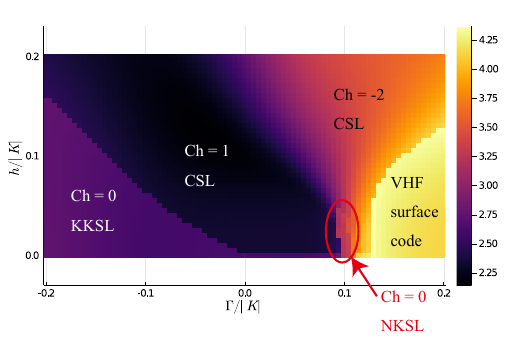}
    \caption{Phase diagram obtained from the 54-site exact
    diagonalization of the K$\Gamma$ model.  The color plot
    shows the value of the entanglement entropy.  The entanglement
    cut is shown in Fig.~\ref{tqs}(a) by the blue dashed circle
    (see also SM~\cite{SM}).}
    \label{see24}
\end{figure}

Finally, we check the value of $V_p$ to identify the mysterious
area which cannot be identified by the calculation of the nematic
order parameter.  As shown in Fig.~\ref{phase}(d), the value of
$|V_p|$ reaches around 0.75, which is very close to the highest value
$+1$.  Combined with what is obtained from the $tqs$ model,
we believe that this phase is connected to the large $|s|$ phase
of the $tqs$ model.  Thus, this phase appearing in the large $\Gamma/|K|$
region should be described by the VHF surface code.  This region
indeed has the largest value of the entanglement entropy (see
Fig.~\ref{see24}), which reconfirms that it is a highly entangled
phase which cannot be described by the mean-field picture.

\textit{Phase diagram of the $K\Gamma$ model}. ---
As shown in Fig.~\ref{see24}, we find the following phases.
\begin{enumerate}
    \item a KKSL with a Chern number $0$.
    \item Kitaev's non-Abelian CSL with a Chern number $1$.
    \item an Abelian CSL with a Chern number $-2$.
    \item an NKSL with a Chern number $0$.
    \item a VHF surface code.
\end{enumerate}
Except for the fact that the zigzag nematic phase is replaced
by the VHF surface code, the phases obtained are the same as
those found in the mean-field theory.
We note that Kitaev's non-Abelian CSL with a Chern number $1$
is identified because this phase is stable on the $\Gamma/|K|=0$ line~\cite{Takahashi2021},
while an Abelian CSL with a Chern number $-2$ is just a speculation.

Especially in the experimentally relevant region, $\Gamma/|K|\sim 0.1$,
we find an NKSL phase with a Chern number 0, which potentially explains
the observed topological nematic transition from Kitaev's non-Abelian
CSL to the toric code phase.

Overall, we have obtained a very rich phase diagram within the flux-free
assumption, and disclosed the existence and properties of NKSL expected
from experiments.  Though our calculation is somewhat
idealized, a more realistic phase diagram should be obtained by
dealing with a $\Gamma^\prime$ term, for example, additionally~\cite{Takikawa2019,Takikawa2020}.
In addition, in the future it is important to check whether
an Abelian CSL with a Chern number $-2$ is connected to the one
discovered previously in the K$\Gamma^\prime$ model~\cite{Takikawa2020}.

\textit{Effect of the Heisenberg term}. ---
Here we will discuss the effect of the Heisenberg term.
With this term the third-order perturbation leads
to the following additional interaction.
\begin{equation}
    H_\textrm{eff}^\prime \propto \frac{h^2 J}{|K+J|^2} \left(\sum_{\langle\!\langle\!\langle ijkl \rangle\!\rangle\!\rangle} c_i c_j c_k c_l \pm \sum_{\langle\!\langle\!\langle ij \rangle\!\rangle\!\rangle^\prime} ic_ic_j \right),
\end{equation}
where $\langle\!\langle\!\langle ijkl \rangle\!\rangle\!\rangle$
represents an zigzag-type interaction for sites $ijkl$, and
$\langle\!\langle\!\langle ij \rangle\!\rangle\!\rangle^\prime$
represents an NNNN hopping which is not a diagonal of hexagons.
This term promotes nematic order, and potentially enhances NKSL.
Thus, the additional small Heisenberg term should be relevant to
the field-induced nematic transition, and will be investigated in
the future.

\textit{Discussion}. ---
We invent a systematic approach to treat non-Kitaev interactions in
a perturbative manner by fully incorporating many-body effects of
Majorana interactions.  We discover the emergence of a Kekul\'e
order in the K$\Gamma$ model for $\Gamma<0$, and a nematic
order for $\Gamma>0$, as well as a more exotic VHF surface code
phase.  We also discover an Abelian CSL with a Chern number -2,
which can be detected by the sign change of the thermal Hall effect
or spin Seebeck effect~\cite{Takikawa2021}.

In the experimentally relevant region $\Gamma/|K| \sim 0.1$ for
$\alpha$-RuCl$_3$, the discovered nematic phase potentially
explains the nematic transition observed in the high-field
phase.  While the position of the NKSL
phase $\Gamma/|K| \sim 0.1$ is consistent the density
matrix renormalization group~\cite{Gohlke2018,Gohlke2020},
its nature is totally different from nematic phases discovered
previously~\cite{Lee2020}. NKSL is not a simple nematic phase
but a spin liquid phase with a partial symmetry breaking.
We believe that NKSL discovered in our study is not the same
phase as the ones in the previous studies, and that the
approximate coincidence of the nematic region
$\Gamma/|K| \sim 0.1$ is only accidental.
Our study endorses the topological nematic transition
scenario proposed in Ref.~\cite{Takahashi2021} with a more
realistic model and interactions.  This state is potentially
detectable by nuclear magnetic resonance (NMR) or M\"ossbauer
spectroscopy~\cite{Yamada2021}.

Another direction to explore is to engineer the Kekul\'e order
in artificial systems.  The sign and magnitude of the $\Gamma$
term is known to be strongly dependent on the spin-orbit
coupling (SOC), so it can potentially be controlled by the proximity
effect with a substrate.  The substrate with heavy elements may
strengthen the SOC of $\alpha$-RuCl$_3$, and allow the fine
tuning of $\Gamma$.  The signature of the Kekul\'e phase is
the existence of an MZM at the $Z_3$ vortex core,
which would potentially be discovered by scanning tunneling
microscope (STM), leading to a possible topological quantum computation.

\begin{acknowledgments}
We thank M.~Gohlke, I.~Kimchi, Y.~Tada, M.~O.~Takahashi, D.~Takikawa
for fruitful discussions.
This work was supported by JSPS KAKENHI Grant No. JP21H01039,
and by JST CREST Grant Number JPMJCR19T5, Japan.
M.G.Y. is supported by Multidisciplinary Research Laboratory System for Future Developments,
Osaka University.
This research was supported in part by the National Science Foundation
under Grant No. NSF PHY-1748958.  The computation in this work has been
done using the facilities of the Supercomputer Center, the Institute
for Solid State Physics, the University of Tokyo.
\end{acknowledgments}

\bibliography{paper}

\end{document}


\title{Supplemental Material for ``Quantum liquid crystals \\ in the finite-field K$\Gamma$ model for $\alpha$-RuCl$_3$''}

\author{Masahiko G. Yamada}
\email[]{myamada@mp.es.osaka-u.ac.jp}
\affiliation{Department of Materials Engineering Science, Osaka University, Toyonaka 560-8531, Japan}
\author{Satoshi Fujimoto}
\affiliation{Department of Materials Engineering Science, Osaka University, Toyonaka 560-8531, Japan}
\affiliation{Center for Quantum Information and Quantum Biology, Osaka University, Toyonaka 560-8531, Japan}

\maketitle

\section{Mean-field approximation}

We here explain the mean-field approximation used
in the main text for the effective models.
It is important to note that the nematic phase and
the Kekul\'e phase cannot be treated by
the same mean-field theory.  Thus, we employed two
mean-field approximations, each of which
is for each bond-ordering pattern.  The discussions
follow Refs.~\cite{Li2018,Takahashi2021}

\begin{figure}
    \centering
    \includegraphics[width=8.6cm]{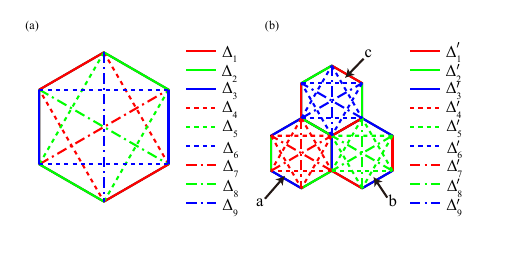}
    \caption{Definition of order parameters.  (a) Order parameters
    $\Delta_\alpha$ for the nematic case.  (b) Order parameters
    $\Delta_\alpha^\prime$ for the Kekul\'e case.  Hexagons are
    labeled $abc$ as indicated by arrows.}
    \label{delta}
\end{figure}

The mean-field approximation is done about order parameters
$\Delta_\alpha$ (and $\Delta_\alpha^\prime$ for the Kekul\'e case)
with $\alpha = 1,\dots,9$ and we assume
$\Delta_\alpha = i\langle c_i c_j \rangle$ to be independent
of a bond $ij$, where a bond $ij$ is defined as shown in
Fig.~\ref{delta} for each $\Delta_\alpha$ (and $\Delta_\alpha^\prime$).
Correspondingly, we can define $\tau_\alpha$ (and
$\tau_\alpha^\prime$ for the Kekul\'e case) with
$\alpha = 1,\dots,9$ as a coefficient of the hopping
term on the $\Delta_\alpha$ bond in the mean-field
Hamiltonian.  We note that all terms are antisymmetrized
in the mean-field Hamiltonian.

\subsection{Nematic phase}

In the nematic case, the mean-field Hamiltonian can be written as
\begin{equation}
    H_\textrm{MF} = \sum_{\langle ij \rangle_\alpha} i\tau_\alpha \eta_{ij} c_i c_j,
\end{equation}
where $\langle ij \rangle_\alpha$ ($\alpha = 1,\dots,9$) means
that a bond $ij$ is connected by the $\Delta_\alpha$-bond,
and $\eta_{ij}=\pm 1$ is defined from the antisymmetrization.
By diagonalizing this one-body Hamiltonian, we can
obtain a band structure.  By filling up the Fermi sea
until half filling, we can obtain the mean-field ground
state $\ket{\Psi_\textrm{MF}}$ and the mean-field energy
$E_\textrm{MF} = \braket{\Psi_\textrm{MF}|H_\textrm{MF}|\Psi_\textrm{MF}}$.
We note that we need a numerical integration about
filled energy bands during the calculation of $E_\textrm{MF}$.

Utilizing the Hellmann-Feynman theorem, we can derive
\begin{align}
    \Delta_\alpha &= \frac{1}{N}\frac{\partial E_\textrm{MF}}{\partial \tau_\alpha} \quad (\alpha = 1,\,2,\,3,\,7,\,8,\,9), \nonumber \\
    \Delta_\alpha &= \frac{1}{2N}\frac{\partial E_\textrm{MF}}{\partial \tau_\alpha} \quad (\alpha = 4,\,5,\,6). \label{eq:nematic}
\end{align}
Partial derivatives are estimated by numerical differentiation.

The rest is to calculate $E_\textrm{tot}=\braket{\Psi_\textrm{MF}|H_c^\prime|\Psi_\textrm{MF}}$.
While the free part can directly be written by $\Delta_\alpha$,
the interaction energy must be evaluated by Wick's theorem.
For example, using an equation like
\begin{align}
i^3\langle c_1 c_2 c_3 c_4 c_5 c_6 \rangle
=& i\langle c_1 c_2 \rangle i \langle c_3 c_4 \rangle i \langle c_5 c_6 \rangle \nonumber \\
& -i\langle c_1 c_2 \rangle i \langle c_3 c_5 \rangle i \langle c_4 c_6 \rangle \nonumber \\
& +i\langle c_1 c_2 \rangle i \langle c_3 c_6 \rangle i \langle c_4 c_5 \rangle \nonumber \\
& -i\langle c_1 c_3 \rangle i \langle c_2 c_4 \rangle i \langle c_5 c_6 \rangle \nonumber \\
& +i\langle c_1 c_3 \rangle i \langle c_2 c_5 \rangle i \langle c_4 c_6 \rangle \nonumber \\
& -i\langle c_1 c_3 \rangle i \langle c_2 c_6 \rangle i \langle c_4 c_5 \rangle \nonumber \\
& +i\langle c_1 c_4 \rangle i \langle c_2 c_3 \rangle i \langle c_5 c_6 \rangle \nonumber \\
& -i\langle c_1 c_4 \rangle i \langle c_2 c_5 \rangle i \langle c_3 c_6 \rangle \nonumber \\
& +i\langle c_1 c_4 \rangle i \langle c_2 c_6 \rangle i \langle c_3 c_5 \rangle \nonumber \\
& -i\langle c_1 c_5 \rangle i \langle c_3 c_4 \rangle i \langle c_2 c_6 \rangle \nonumber \\
& +i\langle c_1 c_5 \rangle i \langle c_3 c_2 \rangle i \langle c_4 c_6 \rangle \nonumber \\
& -i\langle c_1 c_5 \rangle i \langle c_3 c_6 \rangle i \langle c_4 c_2 \rangle \nonumber \\
& +i\langle c_1 c_6 \rangle i \langle c_3 c_4 \rangle i \langle c_2 c_5 \rangle \nonumber \\
& -i\langle c_1 c_6 \rangle i \langle c_3 c_2 \rangle i \langle c_4 c_5 \rangle \nonumber \\
& +i\langle c_1 c_6 \rangle i \langle c_3 c_5 \rangle i \langle c_4 c_2 \rangle,
\end{align}
we can obtain the six-body interaction.  Therefore,
\begin{align}
i^3\langle c_1 c_2 c_3 c_4 c_5 c_6 \rangle
=& -\Delta_1 \Delta_2 \Delta_3
+\Delta_3 \Delta_4 \Delta_5
+\Delta_3^2 \Delta_9 \nonumber \\
& +\Delta_2 \Delta_4 \Delta_6 
+\Delta_4^2 \Delta_7
+\Delta_3 \Delta_4 \Delta_5 \nonumber \\
& +\Delta_2^2 \Delta_8
-\Delta_7 \Delta_8 \Delta_9
+\Delta_5^2 \Delta_8 \nonumber \\
& +\Delta_1 \Delta_5 \Delta_6
+\Delta_2 \Delta_4 \Delta_6
+\Delta_6^2 \Delta_9 \nonumber \\
& +\Delta_1^2 \Delta_7
-\Delta_1 \Delta_2 \Delta_3
+\Delta_1 \Delta_5 \Delta_6 \nonumber \\
=& -2\Delta_1 \Delta_2 \Delta_3 -\Delta_7 \Delta_8 \Delta_9 \nonumber \\
& +2\Delta_3 \Delta_4 \Delta_5 +2\Delta_2 \Delta_4 \Delta_6 + 2\Delta_1 \Delta_5 \Delta_6 \nonumber \\
& +(\Delta_1^2 + \Delta_4^2) \Delta_7 +(\Delta_2^2 + \Delta_5^2) \Delta_8 \nonumber \\
& +(\Delta_3^2 + \Delta_6^2) \Delta_9. 
\end{align}

A similar reduction is possible for other terms and
we can compute $E_\textrm{tot}$.  Self-consistent equations
can be derived by minimize $E_\textrm{tot}$ about $\tau_\alpha$,
and by using Eq.~\eqref{eq:nematic}.  The obtained self-consistent
equations are solved iteratively, while each iteration requires
diagonalization, numerical integration and numerical differentiation.

\subsection{Kekul\'e phase}

In the Kekul\'e case, we consider
\begin{equation}
    H_\textrm{MF}^\prime = \sum_{\langle ij \rangle_\alpha} i\tau_\alpha^\prime \eta_{ij} c_i c_j.
\end{equation}
In the same way as the nematic case, we can calculate
$E_\textrm{MF}^\prime = \braket{\Psi_\textrm{MF}^\prime|H_\textrm{MF}^\prime|\Psi_\textrm{MF}^\prime}$.

From the Hellmann-Feynman theorem,
\begin{align}
    \Delta_\alpha^\prime &= \frac{1}{N}\frac{\partial E_\textrm{MF}^\prime}{\partial \tau_\alpha^\prime} \quad (\alpha = 1,\,2,\,3,\,7,\,8,\,9), \nonumber \\
    \Delta_\alpha^\prime &= \frac{1}{2N}\frac{\partial E_\textrm{MF}^\prime}{\partial \tau_\alpha^\prime} \quad (\alpha = 4,\,5,\,6).
\end{align}

As an illustration, we only compute the six-body
term.  Differently from the nematic case, the results
depend on hexagons.  As for the $a$-hexagon defined in
Fig.~\ref{delta}(b),
\begin{align}
i^3\langle c_1 c_2 c_3 c_4 c_5 c_6 \rangle_a
=&-\Delta_3^{\prime 3} + \Delta_3^{\prime} \Delta_4^{\prime 2} + \Delta_2^{\prime} \Delta_3^{\prime} \Delta_7^{\prime} \nonumber \\
&+\Delta_3^{\prime} \Delta_4^{\prime 2} + \Delta_4^{\prime 2} \Delta_7^{\prime} + \Delta_2^{\prime} \Delta_4^{\prime 2} \nonumber \\
&+\Delta_2^{\prime} \Delta_3^{\prime} \Delta_7^{\prime} - \Delta_7^{\prime 3} + \Delta_4^{\prime 2} \Delta_7^{\prime} \nonumber \\
&+\Delta_3^{\prime} \Delta_4^{\prime 2} + \Delta_2^{\prime} \Delta_4^{\prime 2} + \Delta_4^{\prime 2} \Delta_7^{\prime} \nonumber \\
&+\Delta_2^{\prime} \Delta_3^{\prime} \Delta_7^{\prime} - \Delta_2^{\prime 3} + \Delta_2^{\prime} \Delta_4^{\prime 2} \nonumber \\
=&-\Delta_2^{\prime 3} - \Delta_3^{\prime 3} - \Delta_7^{\prime 3} \nonumber \\
&+3\Delta_4^{\prime 2}(\Delta_2^{\prime} + \Delta_3^{\prime} + \Delta_7^{\prime}) \nonumber \\
&+3\Delta_2^{\prime} \Delta_3^{\prime} \Delta_7^{\prime}.
\end{align}

As for the $b$-hexagon,
\begin{align}
i^3\langle c_1 c_2 c_3 c_4 c_5 c_6 \rangle_b
=&-\Delta_3^{\prime 3} - \Delta_1^{\prime 3} - \Delta_8^{\prime 3} \nonumber \\
&+3\Delta_5^{\prime 2}(\Delta_3^{\prime} + \Delta_1^{\prime} + \Delta_8^{\prime}) \nonumber \\
&+3\Delta_3^{\prime} \Delta_1^{\prime} \Delta_8^{\prime},
\end{align}

Similarly, as for the $c$-hexagon,
\begin{align}
i^3\langle c_1 c_2 c_3 c_4 c_5 c_6 \rangle_c
=&-\Delta_1^{\prime 3} - \Delta_2^{\prime 3} - \Delta_9^{\prime 3} \nonumber \\
&+3\Delta_6^{\prime 2}(\Delta_1^{\prime} + \Delta_2^{\prime} + \Delta_9^{\prime}) \nonumber \\
&+3\Delta_1^{\prime} \Delta_2^{\prime} \Delta_9^{\prime}.
\end{align}

By summing up all these terms, we can obtain the expectation
value of six-body terms.  The derivation and calculation of
self-consistent equations are the same as those for the nematic
case.

\section{Gap opening}

\begin{figure}
    \centering
    \includegraphics[width=8cm]{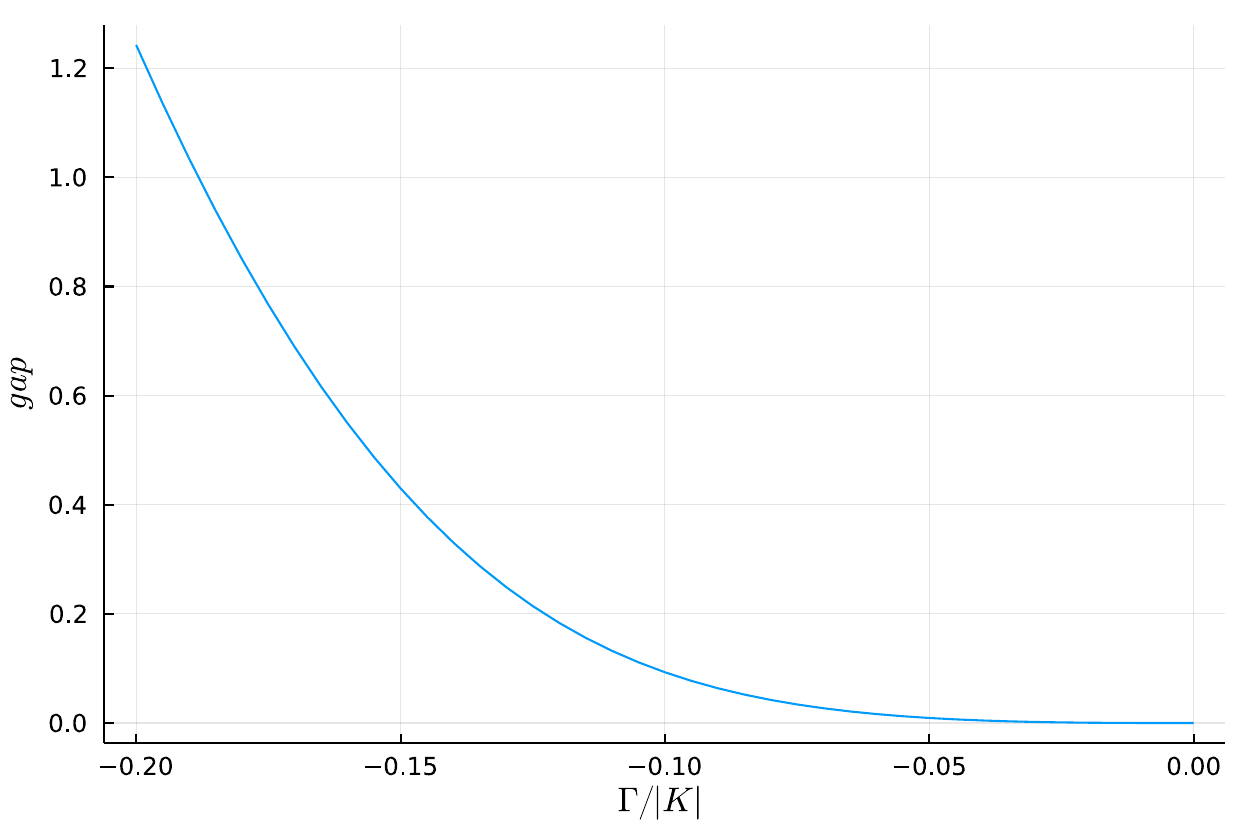}
    \caption{Gap opening by exact diagonalization.}
    \label{gap}
\end{figure}

As stated in the main text, Kitaev spin liquid may be
unstable with respect to a negative $\Gamma$ perturbatively
and it is possible that the Kekul\'e order opens a gap for
Dirac cones. In order to see this we checked the gap opening
within the exact diagonalization.  On the $h/|K|=0$ line,
the gap opens from $\Gamma/|K|=0$ to $\Gamma/|K|=-0.2$,
as shown in Fig.~\ref{gap}.
The original fourfold degeneracy is lifted by an infinitesimal
$\Gamma/|K|<0$.

\section{Entanglement entropy}

\begin{figure}
    \centering
    \includegraphics[width=4cm]{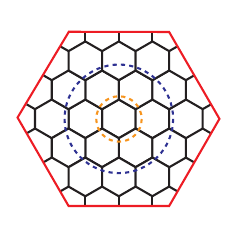}
    \caption{Entanglement cuts.}
    \label{ee}
\end{figure}

\begin{figure}
    \centering
    \includegraphics[width=8cm]{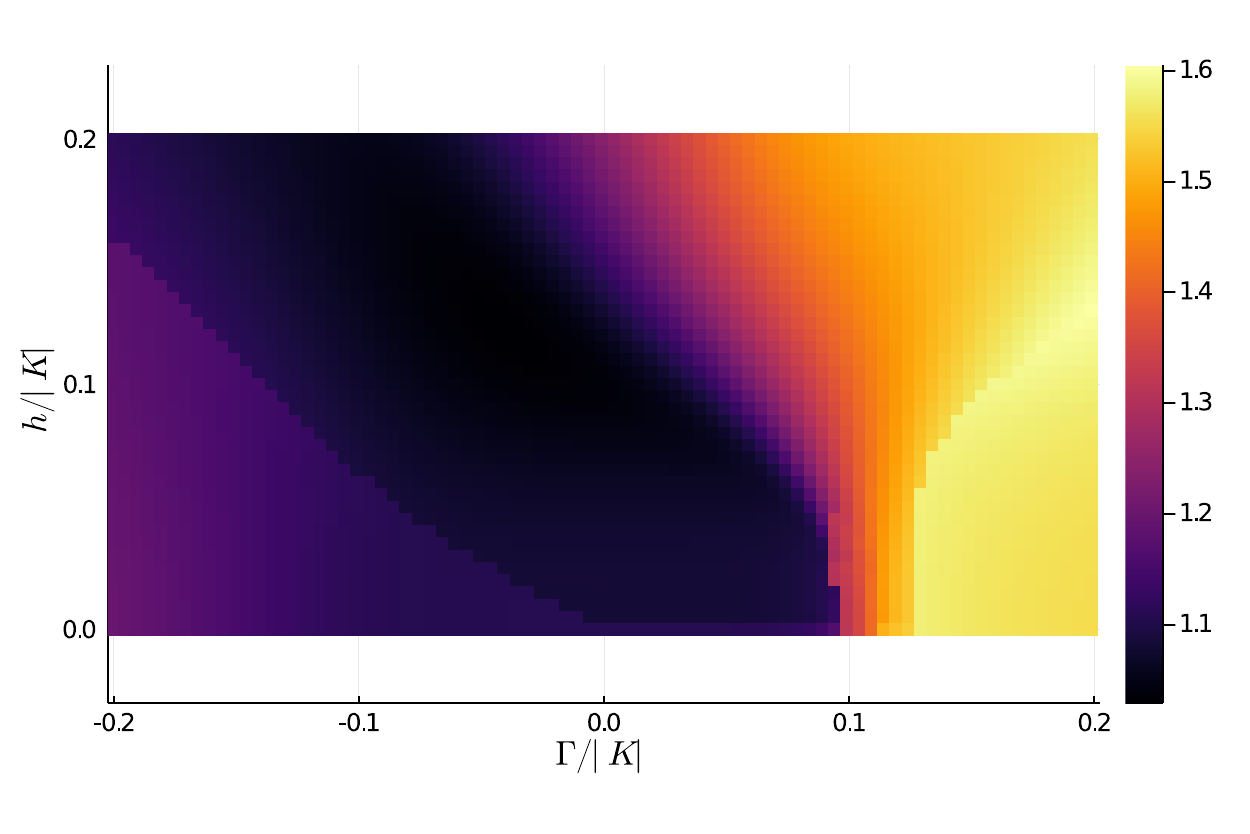}
    \caption{Entanglement entropy calculated by the 6-site
    cut for the 54-site system.}
    \label{see6}
\end{figure}

The calculation of entanglement entropy is done by decomposing
the Hilbert space as usual.  We must be careful about the
decomposition because in Majorana systems we cannot cut between
a pair of sites, where two Majorana fermions are combined into
a complex fermion.

In the 54-site calculation, the entanglement cut is
done by 6 sites (orange dashed circle) and 24 sites
(blue dashed circle) as shown in Fig.~\ref{ee}.
The entanglement entropy calculated by the 24-site cut
is shown in Fig.~4 in the main text.  The entanglement
entropy calculated by the 6-site cut is shown in Fig.~\ref{see6}.

\bibliography{suppl}